\documentclass[prd,aps,preprintnumbers,twocolumn]{revtex4}
\usepackage{epsfig}
\usepackage{amsmath}

\begin{document}

\preprint{BNL-NT-06/20} 
\preprint{YITP-SB-06-18}

\def\slash#1{\ooalign{$\hfil/\hfil$\crcr$#1$}} 

\renewcommand{\thefigure}{\arabic{figure}}

\newcommand\bea{\begin{eqnarray}}
\newcommand\eea{\end{eqnarray}}

\def\ltap{\raisebox{-.4ex}{\rlap{$\,\sim\,$}} \raisebox{.4ex}{$\,<\,$}}
\def\gtap{\raisebox{-.4ex}{\rlap{$\,\sim\,$}} \raisebox{.4ex}{$\,>\,$}}
\def\lra{\leftrightarrow}
\def\naive{na\"{\i}ve}
\newcommand\as{\alpha_s}
\newcommand\f[2]{\frac{#1}{#2}}
\def\la{\lambda} 
\def\LN{\ln N} 
\def\beq{\begin{equation}}
\def\eeq{\end{equation}}
\def\beeq{\begin{eqnarray}}
\def\eeeq{\end{eqnarray}}
\def\to{\rightarrow}
\def\nn{\nonumber}
\def\arrowlimit#1{\mathrel{\mathop{\longrightarrow}\limits_{#1}}}
\def\qt{q_{\perp}}
\def\res{{\rm res.}}
\def\ms{${\overline {\rm MS}}$} 
\def\msbar{{\overline {\rm MS}}} 
\def\asp{{\alpha_s}\over{\pi}} 
\def\b0{b_0}
\def\bone{b_1}
\def\btwo{b_2}
\def\GE{\gamma_E}
\newcommand\vep{\varepsilon}

\title{Crossed Threshold Resummation}

\author{George Sterman$^1$, Werner Vogelsang$^2$}

\affiliation{${}^1$C.N.\ Yang Institute for Theoretical Physics,
Stony Brook University 
Stony Brook, New York 11794 -- 3840, U.S.A. \\
${}^2$Physics Department, 
Brookhaven National Laboratory,
Upton, New York 11973, U.S.A.}
\date{\today}

\begin{abstract}
We show that
certain general properties of threshold and joint resummations in 
Drell-Yan
cross sections hold as well for their crossed analogs in semi-inclusive
deep-inelastic scattering and double-inclusive leptonic
annihilation. We show that all plus-distribution corrections
near threshold show the same structure, and are determined
to all logarithmic order by
two anomalous dimensions, one of which is a generalization
of the $D$-term previously derived in Drell-Yan.
We also discuss the possibility of universality in
power corrections implied by the resummation.
\end{abstract}

\maketitle

\section{Introduction}

Resummations  organize sets of potentially large higher-order contributions
to perturbative series.
In hard inclusive and semi-inclusive cross sections,
such corrections typically arise when restrictions
on phase space result in incomplete 
cancellations between partonic emission and
virtual corrections.  The realization of a cross section 
that respects unitarity may then require the resummation 
of a portion of the infinite series, involving
unlimited numbers of partons.

  For hadron-hadron scattering processes, both
transverse momentum~\cite{Working02,qtresum}
and threshold~\cite{dyresum1,dyresum2} resummations have been 
studied extensively, especially for differential
and inclusive cross sections for
electroweak vector and heavy scalar (Higgs) production.  
In these cases, the technique 
of joint resummation~\cite{LSV,KSV,Banfi04} provides a unification
of the two methods in situations where both
may be important.    In addition, transverse momentum
resummation has also found applications in 
crossed processes involving the action of electromagnetic
currents, especially doubly-inclusive hadron production
in leptonic annihilation and single-inclusive
hadron production in deep-inelastic scattering~\cite{qtresum1}.

In this short paper, we show how in these cases
there are natural crossed analogs of threshold and joint 
resummations for Drell-Yan (DY) cross sections. Extensions of
this type for threshold resummation 
were previously discussed 
in Ref.~\cite{LSV,cacc}. Here we will introduce specific new
observables in single-inclusive deep-inelastic scattering (SIDIS) 
and double-inclusive leptonic annihilation (DIA).
These are related to the
doubly-differential distributions already studied 
at next-to-leading logarithm in
\cite{cacc}.   We will see that the
singly-differential threshold-resummed cross sections
in this set share a transparent
structure that extends to all logarithms.

These considerations also have implications for
nonperturbative power corrections in these cross sections,
through the running of the QCD coupling \cite{BBpowerreview}.
In principle, perturbation theory applies to all 
infrared safe observables, 
which depend only on large dimensional scales, $Q$, up to
corrections that are formally suppressed by powers of $Q$.
There is a close relationship between 
resummed perturbative predictions
and power corrections, however,
because the perturbative series does not converge.
In our analysis of
observables related by crossing we will propose 
a universal structure for both resummed 
logarithmic and power corrections.
Such universality conjectures on power corrections 
have previously been tested 
extensively in average and differential jet event shapes
\cite{BBpowerreview,DSeventreview} in $\rm e^+e^-$ 
annihilation cross sections, with phenomenological
success. 

In Sec.~\ref{sec2} we introduce the cross sections in DY, DIA,
and SIDIS that we will consider in this paper. 
Section~\ref{sec3} analyzes their underlying partonic cross
sections near threshold, and in  Sections~\ref{sec4} and~\ref{sec5}
we investigate the corresponding threshold resummation. In 
Sec.~\ref{sec6} we discuss the universality properties of the
cross sections, and we extend our
results to the case of joint resummation. 

\section{Crossed Threshold Variables \label{sec2}}

We consider processes that are characterized by the
lowest-order (LO) partonic reaction $q\bar{q}\to\gamma^*$ 
or one of its crossed versions. These are
the Drell-Yan (DY) process $h_1 h_2\to \ell^+\ell^- X$, 
``semi-inclusive'' deeply-inelastic scattering (SIDIS)
$\ell h_1\to \ell h_2X$, and ``double-inclusive leptonic annihilation''
(DIA), $\ell^+\ell^-\to h_1 h_2 X$, where 
$h_{1,2}$ denote hadrons. For each of these processes, 
we are interested in the cross section with the least restrictive kinematics,
and the crossed versions of the variable that controls threshold logarithms
for DY cross sections.

In the DY process, the most inclusive observable
is the usual ``total'' cross section,
differential only in the variable
\beq
\tau_{\rm DY} \equiv \frac{Q^2}{(P_1+P_2)^2}\; ,
\eeq
where $Q$ is the invariant mass of the lepton pair, and $P_{1,2}$
are the momenta of the initial hadrons, so that the denominator is 
the hadronic center of mass energy squared. Cross sections for SIDIS 
and DIA have customarily been considered differential in {\it two}
light-cone scaling variables, one associated with each of the two
hadrons~\cite{aemp,cacc,dfv}. For example, in the case of SIDIS,
one usually employs the Bjorken variable $x\equiv Q^2/2P_1\cdot q$
(with $q$ the momentum of the virtual photon; $Q^2\equiv -q^2$)
and the ``fragmentation'' variable $z\equiv P_1\cdot P_2/P_1\cdot q$
and studies the cross section differential in both of these \cite{cacc}. This
treatment is also followed in 
experimental studies~\cite{hermes}. 
In contrast to this we define, in analogy with the DY process, 
the ``$\tau$-variable''
\beq
\tau_{\rm SIDIS} \equiv xz = 
\left(\frac{Q^2}{2P_1\cdot q}\right)\, 
\left(\frac{P_1\cdot P_2}{P_1\cdot q}\right) \; ,
\eeq
and consider the cross section differential in $\tau_{\rm SIDIS}$. We
are not aware of a discussion of this cross section in the 
earlier literature. As we shall show below, this cross section
has remarkable similarities with the DY cross section. Given the 
experimental studies of SIDIS in terms of $x$ and $z$, it should 
actually be relatively straightforward to perform measurements of the 
cross section differential in the variable $\tau_{\rm SIDIS}$ as well.

Finally, for DIA, we follow~\cite{aemp} to define 
$x\equiv 2P_1\cdot q/Q^2$, and $z\equiv  P_1\cdot P_2/P_1\cdot q$
and, following the same logic as above, we shall consider the cross section 
differential in 
\beq
\tau_{\rm DIA} \equiv xz = 
\left(\frac{2P_1\cdot q}{Q^2}\right)\, 
\left(\frac{P_1\cdot P_2}{P_1\cdot q}\right) 
= \frac{(P_1 + P_2)^2}{Q^2} \; ,
\eeq
where for the last equality we have neglected the
masses of the produced hadrons to introduce their pair invariant 
mass $(P_1 + P_2)^2$. 

Each of the cross sections discussed above is given by a
convolution of parton distribution functions $f_i^h(\xi,\mu)$ in hadron $h$
and/or hadron fragmentation functions $D_i^h(\xi,\mu)$ with partonic 
hard-scattering functions, where $i$ runs over 
quarks, antiquarks, and gluons, and $\mu$ is a factorization scale. 
We shall collectively refer to the parton distributions and
fragmentation functions as ${\cal F}_i^h(\xi,\mu)$. The functions depend 
on the variable 
$\xi$, which for the parton distributions is the light-cone momentum 
fraction of the initial hadron momentum taken by the parton, while for 
the fragmentation functions it is the momentum fraction that
the produced hadron takes from the parent parton. Denoting
parton momenta by $p$ and hadron momenta by $P$, 
we have $p=\xi P$ for
the initial partons and $p=P/\xi$ for the final-state ones. 
The partonic hard-scattering functions 
will be denoted by $\omega^{ij}_{\rm A}$, where again $i,j$ run over 
parton types, and ${\rm A}={\rm DY,SIDIS,DIA}$. The 
$\omega^{ij}_{\rm A}$ are perturbative; they begin at lowest order 
with the simple processes $q\bar{q}\to\gamma^*$ (or crossed).
For each of the processes we consider, we introduce a partonic 
$\tau$-variable in terms of the corresponding partonic momenta:
\bea
\hat{\tau}_{\rm DY} &\equiv& \frac{Q^2}{(p_1+p_2)^2}=
\frac{\tau_{\rm DY}}{\xi_1\xi_2} \, ,
\nonumber\\
\hat{\tau}_{\rm SIDIS} &\equiv& 
\left(\frac{Q^2}{2p_1\cdot q}\right)\, 
\left(\frac{p_1\cdot p_2}{p_1\cdot q}\right) =
\frac{\tau_{\rm SIDIS}}{\xi_1\xi_2} \, , \nonumber\\
\hat{\tau}_{\rm DIA} &\equiv& 
\left(\frac{2p_1\cdot q}{Q^2}\right)\, 
\left(\frac{p_1\cdot p_2}{p_1\cdot q}\right) 
=\frac{\tau_{\rm DIA}}{\xi_1\xi_2} \; .
\eea
Apart from dependence on the strong coupling $\alpha_s(\mu)$ and 
on the ratio $Q/\mu$, where $\mu$ is the renormalization/factorization scale, 
the $\omega^{ij}_{\alpha}$ 
will be functions only of the variable $\hat{\tau}_{\rm A}$. The LO 
partonic cross section is in each case proportional to 
$\delta (1-\hat{\tau}_{\rm A})$. From now on, we will choose $\mu=Q$ 
throughout.

In the following, we shall mostly be interested in the behavior 
of the partonic cross sections at large $\hat{\tau}_{\rm A}$, 
$\hat{\tau}_{\rm A}\to 1$. 
Since the parton distributions and fragmentation functions
are steeply falling functions of the $\xi$, they emphasize
the region $\xi_1 \xi_2\sim \tau_{\rm A}$, and therefore 
$\hat{\tau}_{\rm A} \sim 1$. For $\hat{\tau}_{\rm A}\to 1$,
the partonic cross sections at higher orders in perturbation theory
develop large logarithmic corrections. For example, for the 
Drell-Yan process, the emission of a single gluon in the
process $q\bar{q}\to\gamma^*$ gives rise to a leading term 
of the form $\alpha_s\left[ \ln(1-\hat{\tau}_{\rm DY})/
(1-\hat{\tau}_{\rm DY})\right]_+$, where the plus-distribution
makes the cross section integrable at $\hat{\tau}_{\rm DY}= 1$ in
the standard way. At yet higher orders in $\alpha_s$, one finds
leading terms of the form $\alpha_s^k\left[ \ln^{2k-1}
(1-\hat{\tau}_{\rm DY})/(1-\hat{\tau}_{\rm DY})\right]_+$, plus
subleading terms that are down by one or more powers of the
logarithm. 

As we shall discuss below, the logarithms arising at 
$\hat{\tau}_{\rm A}\to 1$ have a universal structure in the three cross
sections we are discussing in this paper. They are associated with 
emission of relatively soft gluons into the final state. The analysis
of the large corrections is most conveniently performed in a
reference frame where the {\it energy} of the emitted soft gluons 
is the only relevant quantity. For the Drell-Yan process, this approach 
has been followed in earlier treatments~\cite{dyresum1}, where 
the resummation of the large logarithms to all orders of perturbation 
theory was derived. The frame chosen here is the center-of-mass
frame of the initial hadrons. The large corrections then arise
when the partons have ``just enough'' energy to produce the
final state. For this reason, the logarithms are also referred 
to as threshold logarithms, and their resummation as threshold
resummation.  

Restricting ourselves to the terms in the partonic cross sections
that dominate at large $\hat{\tau}_{\rm A}$, we find the
following generic structure for our three cross sections
of interest:
\bea
\frac{d\sigma_{\rm A}(\tau_{\rm A})}{d\tau_{\rm A}}&=&
\sigma^0_{\rm A}\sum_{i=q,\bar{q}}
\int_{\tau_{\rm A}}^1 \frac{d\xi_1}{\xi_1}{\cal F}_i^{h_1}(\xi_1,Q)
\nonumber 
\\
&&\hspace*{-15mm}\times\,\int_{\tau_{\rm A}/\xi_1}^1 \frac{d\xi_2}{\xi_2}
{\cal F}_i^{h_2}(\xi_2,Q)\,
\omega^i_{\rm A}(\hat{\tau}_{\rm A},\alpha_s(Q))\; ,
\label{convol}
\eea
where for ${\rm A}={\rm DY}$ ${\cal F}_i^{h_1}\equiv 
f_i^{h_1}$, ${\cal F}_i^{h_2}\equiv f_{\bar{i}}^{h_2}$, 
for ${\rm A}={\rm SIDIS}$ ${\cal F}_i^{h_1}\equiv f_i^{h_1}$,
${\cal F}_i^{h_2}\equiv D_i^{h_2}$, and for ${\rm A}={\rm DIA}$
${\cal F}_i^{h_1}\equiv D_i^{h_1}$, ${\cal F}_i^{h_2}\equiv 
D_{\bar{i}}^{h_2}$. The normalization $\sigma^0_{\rm A}$ 
is specific to each process; it may depend on additional 
variables such as lepton scattering angles (the cross section
may also be differential in these, in addition to $\tau_{\rm A}$).
It is chosen in such a way that each of the $\omega^i_{\rm A}$
begins with $\delta (1-\hat{\tau}_{\rm A})$ at LO.
Note that the sum in Eq.~(\ref{convol}) only runs over quarks and
antiquarks and is diagonal in flavor. This is because soft gluon emission 
gives rise to  threshold logarithms only
in the process $q\bar{q}\to\gamma^*$ (or crossed). 
Partonic channels with an initial gluon, or
with a gluon fragmenting into the observed hadron, are suppressed
near threshold.   For this reason, we keep only a single 
partonic index on $\omega_A$.

Also note that in general there could be two
terms of the form in~(\ref{convol}) for a given process, depending on 
whether the virtual photon carries transverse or longitudinal polarization. 
For example, in DIS, there are contributions involving the structure
functions $F_1$ and $F_L$. However, it turns out that near partonic 
threshold the longitudinal polarization component is suppressed, so that 
there is only a single structure like Eq.~(\ref{convol}) for each
process. 

In the following, it will be convenient to introduce Mellin moments
of the cross sections in Eq.~(\ref{convol}) in $\tau_{\rm A}$. 
Defining for any function $f(x)$ the moments $\tilde{f}(N)\equiv \int_0^1 
dx x^{N-1}f(x)$, we find:
\bea
\hspace*{-3mm}\tilde{\sigma}_{\rm A}(N)&=&
\sigma^0_{\rm A}\sum_{i=q,\bar{q}}
\tilde{{\cal F}}_i^{h_1}(N,Q)\tilde{{\cal F}}_i^{h_2}(N,Q)\,
\tilde{\omega}^i_{\rm A}(N,\alpha_s(Q)) \; .\nonumber \\ 
&&
\label{moments}
\eea
Small $1-\hat{\tau}_{\rm A}$ corresponds to large $N$. The  
inverse transformation reads:
\bea
{d\sigma_A \over d\tau_A}
&=& \sigma^0_A\; \sum_{i=q,\bar{q}}\;
\int_{\cal C} {dN \over 2 \pi i}\; \tau_A^{-N}\,
\tilde{\cal F}_i^{h_1}(N,Q) \tilde{\cal F}_i^{h_2}(N,Q)\;
\nonumber\\
&& \hspace{5mm} \times\ \tilde \omega^i_A(N,\alpha_s(Q)) \, ,
\label{physsig}
\eea
where ${\cal C}$ denotes a contour in complex-$N$ space.
In moment space, the partonic hard-scattering functions 
have the perturbative expansion
\beq
\tilde\omega^i_{\rm A}(N,\alpha_s(Q))=1+\frac{\alpha_s(Q)}{2\pi}
\tilde\omega^{i(1)}_{\rm A}(N)+{\cal O}(\alpha_s(Q)^2)\; .
\eeq
The explicit forms of the coefficients $\tilde
\omega^{i(1)}_{\rm A}(N)$
may be obtained from results in the literature~\cite{aemp,aem,dfsv}. 
One finds for large $N$:
\bea
\tilde\omega^{i(1)}_{\rm DY}(N)&=&4 \ln^2(\bar{N})
-8+\frac{2}{3}\pi^2\,+\,{\cal O}
(1/N)\nonumber \\
&=&
\tilde\omega^{i(1)}_{\rm DIA}(N) \; ,\nonumber \\[2mm]
\tilde\omega^{i(1)}_{\rm SIDIS}(N)&=&\tilde
\omega^{i(1)}_{\rm DY}(N)-\pi^2\,+\,{\cal O}(1/N)\; ,
\label{oneloop}
\eea
where $\bar{N}\equiv N{\mathrm{e}}^{\gamma_E}$ with $\gamma_E$ the Euler
constant. The logarithmic term is the moment-space equivalent of 
the threshold logarithm $\left[ \ln(1-\hat{\tau}_{\rm A})/
(1-\hat{\tau}_{\rm A})\right]_+$ in $\hat{\tau}_{\rm A}$ space
mentioned above. As we anticipated, this term is universal in all
$\omega^{i(1)}_{\rm A}(N)$. The $N$-independent pieces partly result 
from virtual corrections, which explains the difference $\sim \pi^2$
between the time-like DY and DIA processes and the space-like 
SIDIS.   We will show below that the close relationship 
between the $\tilde{\omega}_A^i$ extends to all orders.

\section{Phase Space near Partonic Threshold \label{sec3}}

The one-loop double logarithmic structure
 in the moment variable $N$ 
exhibited in the previous subsection can be
generalized and resummed to all orders in
each of the DY, SIDIS and DIA processes. 
The same reasoning will enable us to exhibit
jointly resummed cross sections that organize
the singular behavior in both $N$ and the
impact parameter $b$ conjugate to
the transverse momentum of soft gluon radiation.
The key to these results is an analysis of
the phase space available to partonic radiation
near threshold in each case.   We will find that,
by a suitable choice of frame, all logarithmic
behavior in the partonic variables  $\tau_A$,
$A=$ DY, SIDIS or DIA, arises from a
region where there is a  limitation on
the {\it total energy} of partonic radiation.
The powerful consequences of this restriction
were explored in detail for Drell-Yan processes
in Ref.\ \cite{LSV}.   We derive the
analogous results for SIDIS and DIA
in the following section.    
Here, we review the DY phase
space and provide its extensions to the other cases.

For Drell-Yan processes $h_1(P_1) h_2(P_2)\to \ell^+\ell^- X$, 
we choose the overall partonic rest frame, with 
\bea
(p_1 + p_2)^\mu = \sqrt{\hat s}\delta_{\mu0}\, .
\eea
The relation between the observed vector boson momentum $q$ and
the momentum $k$ of unobserved radiation is
\bea
p_1+p_2 = q + k\, .
\eea
In the rest frame, we then evaluate the partonic $\hat \tau_A$ variable 
and find
\bea
\hat \tau_{\rm DY} = \frac{(p_1+p_2 - k)^2}{\hat s} = 1  - 
\frac{2k^0}{Q} + {\cal O}\left[ (1-\hat \tau_{\rm DY})^2\right] .
\label{dyk0}
\eea
Thus, the difference between $\hat \tau_{\rm DY}$ 
and unity is twice the total energy of partonic
radiation divided by $Q$, up to corrections that
vanish as a power for $\hat \tau_{\rm DY} \to 1$, and which
therefore do not affect logarithmic behavior in the moments.

To construct a similar analysis for the SIDIS
process $\ell h_1(P_1)\to \ell h_2(P_2)X$, 
we choose a partonic Breit frame where 
\bea
&& \hspace{10mm} p_1= 
(p_1^+,0^-,0_T)\, , 
\nonumber\\
&& \hspace{10mm}  q = (- q^+, Q^2/2q^+,0_T )\, ,
\nonumber\\
&& \hspace{10mm} p_1^+=q^- \, .
\eea
As before, we define $k^\mu$ as the momentum of all unobserved radiation, 
\bea
p_1+q = p_2 + k\, .
\eea
We then have $k_T = - p_{2,T}$ for the transverse components. 
A brief calculation gives
\bea
\hat \tau_{\rm SIDIS} &=& \left(1 - \frac{k_T^2}{2p_1^+q^-} - 
\frac{k^+}{p_1^+}\right)\, 
\left(1 - \frac{k^-}{q^-}\right) \nonumber \\
&\sim& 1 - \frac{2k^0}{Q} + {\cal O}\left[ 
(1-\hat{\tau}_{\rm SIDIS})^2\right] \, ,
\label{disk0}
\eea
a result precisely analogous to (\ref{dyk0}),
up to nonsingular corrections.

Finally, for DIA $\ell^+\ell^-\to h_1(P_1) h_2(P_2) X$, 
we use the overall rest frame, with
\bea
q^\mu = Q\delta_{\mu 0}\, .
\eea
The relation between the observed momenta and
inclusive radiation in this case is
\bea
q = p_1 + p_2 +k\, ,
\eea
and we again find:
\bea
\hat{\tau}_{\rm DIA} = \frac{(q-k)^2}{2P_1\cdot q} = 
1 - \frac{2k^0}{Q} + {\cal O}\left[ (1-\hat{\tau}_{\rm DIA})^2\right] \, .
\label{diak0}
\eea
Thus, in each case the available phase space
is most easily characterized by a limitation
on the total energy of gluon radiation.  

\section{Threshold resummation \label{sec4}}

Near partonic threshold, in each of the cross
sections discussed above singular behavior
appears as plus-distributions, up to
$[\ln^{2n-1}(1-\hat{\tau}_A)/(1-\hat{\tau}_A)]_+$ in
momentum space and $\ln^{2n}N$ in Mellin-moment 
space at $n$th order in $\as$, with
$A$ = DY, SIDIS, DIA.   Using the results
of the previous section, we can write a
universal form for threshold resummation
in these processes.

As discussed in some detail in Ref.\ \cite{LSV},
all singular corrections in the region near partonic 
threshold for the Drell-Yan process can be factorized
into parton distribution functions in convolution with 
an inclusive cross section for the production of soft 
radiation with total momentum $k$. In the appropriate frame, 
corrections to this factorization are suppressed by powers of the
energy, $k_0/Q$ to the hard-scattering scale $Q$, in momentum
space, and by powers of the moment variable $N$
in the transform space. 

Because we have established that the threshold regions in SIDIS and DIA
are also characterized by small energy of soft gluon radiation $k_0$,
essentially identical arguments apply in these cases as well.  The only 
significant difference is that, as  in Eq.\ (\ref{convol}),
for SIDIS one of the parton distributions is replaced in the convolution by 
a fragmentation function, while for DIA both are replaced.  

For soft emission, the hard scattering functions
$\tilde{\omega}^i_A$
may be evaluated in the eikonal approximation for the
quarks and/or antiquarks involved in the hard scattering.
The fermions are then represented by recoilless
color sources, characterized by velocities $\beta$ and
$\beta'$ in opposite directions along the axis defined
by the relevant
center-of-mass frames.  In these frames, the 
$\tilde{\omega}^i_A$ are invariant under
both boosts and rescalings of the eikonal velocities
$\beta,\beta'$. In moment space, the eikonal hard
scattering functions exponentiate~\cite{LSV,webs}:
\begin{widetext}
\beeq
\tilde \omega^i_A(N,Q) &=& \exp\, \Bigg(\,
2\, \int_{k_T^2<Q^2} {d^{2-2\epsilon}k_T\over 
[2\pi^{1-\varepsilon}/\Gamma(1-\varepsilon)]}\ \Bigg\{\int_0^{Q^2-k_T^2} dk^2\;
W^i_A \left(k_T^2,k_T^2+k^2,\mu^2,\alpha_s(\mu,\varepsilon),
\varepsilon\right)
\nonumber \\[2mm]
& & \hspace{5mm} \times
\left[\, 
K_0\left(2N\sqrt{k_T^2+k^2\over Q^2} \right) - \ln\sqrt{Q^2\over
k_T^2+k^2}\, \right]+ {2\over (k_T^2)^{1-\varepsilon}}\ln \bar N\, 
A_i\left(\as(k_T,\varepsilon)\right) \Bigg\}\, \Bigg)\, ,
\label{nextA2}
\eeeq
\end{widetext}
where we have chosen dimensional regularization with $d=4-2
\varepsilon$ dimensions, in order to make all contributions
to the exponent individually finite.  We have suppressed
the finite dependence on $\varepsilon$ in the function $\tilde{\omega}^i_A$.
All dynamical information in Eq.\ (\ref{nextA2}) is 
contained in the ``web"~\cite{webs} functions 
$W^i_A \left(k_T^2,k_T^2+k^2,\mu^2,\alpha_s(\mu,\vep),\vep \right)$.
The web functions possess the same
boost-invariance and scaling properties as the full eikonal
cross sections.  As such, they can depend only on the boost
invariant quantities $k^2$ and $k_T^2$, the squared total invariant mass
and transverse momentum of radiation.
In the frame we have chosen, $k^2+k_T^2 = k\cdot \beta \,
k\cdot \beta'/(\beta\cdot\beta')$.  
The web functions may also
depend on the boost-invariant signs: ${\rm sgn}(\beta'\cdot k)$, 
which in general leads to differences between the web
functions for the various processes at a non-leading level.
An example at one loop is the $\pi^2$ term in Eq.\ (\ref{oneloop}).

 Webs can be defined graphically in terms of sums of diagrams that are
irreducible by cutting two eikonal lines, and the web functions
are the result of summing over all final-state cuts 
of these diagrams, integrating
over phase space at fixed total final-state momentum, $k$.
The web functions for the three cross sections under consideration
are all related simply by crossings of the eikonal lines.

The boost invariance of the webs allows us to integrate 
over one light-cone component
of $k$, resulting in the Bessel function $K_0$ in Eq.\ (\ref{nextA2}),
with the specific boost-invariant momentum dependence shown.
Corrections to (\ref{nextA2}) are exponentially suppressed in $N$
for all contributions with radiation.

The logarithmic term in the square brackets accounts for
virtual corrections, that is, final states without radiation.  Only these 
contributions are sensitive to the upper limits of the $k$ integrals
for large $N$.  We will see that  in (\ref{nextA2}) the
virtual corrections are {\it defined} to set $\tilde\omega^i_A$ 
to zero at $N=0$.  This condition, of course, does not affect the
logarithmic large-$N$ behavior in moment space, or the singular $\tau_a$
behavior in momentum-space convolutions.
 
The final term in (\ref{nextA2}) represents
the subtraction of collinear singularities. In this term, the
anomalous dimension $A_i(\as)$ is the same coefficient
that appears in the single-log term in the moments of the 
$i\to i$ splitting function, $\tilde P_{ii}(\as,N)
= -A_i(\as)\ln \bar{N} + \dots $ \cite{LSV}. $A_i(\as)$ is 
the same for parton distributions and for fragmentation
functions because it is determined by the elastic form factor.
$i$ and $\bar{i}$, which are both either a quark or an antiquark, 
have the same subtraction of collinear singularities, hence the
factor 2 in this term  of Eq.\ (\ref{nextA2}).

Finally, we recall the very useful property that the web 
functions are invariant under variations
of the renormalization scale~\cite{LSV}:
\bea
\frac{\partial}{\partial \ln \mu^2} 
W^i_A \left(k_T^2,k_T^2+k^2,\mu^2,\alpha_s(\mu,\vep),\vep 
\right) = 0\, .
\label{rginvce}
\eea
This enables us to shift the scale of the running coupling with
the integration over momentum, a feature that we will 
exploit in the next section.

\section{Leading and Subleading Logarithms in the 
Resummed Cross Sections \label{sec5}}

In this section we will provide a new analysis of the exponents
of the resummed cross sections in moment space, Eq.\ (\ref{nextA2}),
that will for the first time relate the exponent in terms of
webs to the anomalous dimensions that have customarily been
introduced to generate nonleading logarithms.  These have variously been 
denoted by $g_3(\as)$ \cite{dyresum1} and 
$D(\as)$ \cite{Drefs}, and are process dependent.  We will see that
for the set of cross sections discussed here, these anomalous
dimensions are very closely related.

In the double integral of Eq.\ (\ref{nextA2}), we change variables from $k^2$ 
to
\vspace*{-0.1cm}
\bea
u^2\equiv k^2 + k_T^2\, ,
\eea
and in the  integral over the term proportional to $A_i$ 
we separate the angular integration and relabel the 
radial  variable $k_T^2$ as $u^2$.   Using the renormalization scale
invariance of the web functions, Eq.~(\ref{rginvce}), 
we choose $\mu=u$ and write
the logarithm of the eikonal cross section as
\begin{widetext}
\beeq
\ln\; \tilde \omega^i_A(N,Q)
&=&
\int_0^{Q^2} du^2\; \Bigg\{\int_0^{u^2} \, dk_T^2 \,
(k_T^2)^{-\varepsilon} \, 
W^i_A \left(k_T^2,u^2,u^2,\alpha_s(u,\varepsilon),
\varepsilon\right)
\nonumber \\
& & \hspace{5mm} \times
\left[\, 
K_0\left({2Nu\over Q} \right) - \ln{Q\over
u}\, \right]\
+ \frac{2}{u^2}\ln \bar N\, 
A_i\left(\as(u,\varepsilon )\right) \Bigg\} \, .
\label{nextAu}
\eeeq
\end{widetext}
In the first term of
this expression, all $k_T$-dependence is in the web function at 
fixed $u$,  and it is natural to define
an anomalous dimension that is a function of $u$ only:
\bea
\frac{\rho^i_A \left (\as(u,\vep),\vep 
\right)}{u^2} &\equiv& \int_0^{u^2} 
dk_T^2 \,
(k_T^2)^{-\varepsilon} \, \nonumber \\
&&\hspace*{-20mm}
\times\;W^i_A \left(k_T^2,u^2,u^2,\alpha_s(u,\varepsilon),
\varepsilon\right) \, .
\eea
The function $\rho^i_A$ defined in this fashion is
dimensionless, so that the overall factor $u^{-2}$ carries
all dimensional information.  Note that
in the subtraction term of Eq.\ (\ref{nextAu}) the power
is also $u^{-2}$ with no dependence on $\varepsilon$.
These integrals are defined by reexpanding the
coupling $\as(u,\vep)$ in terms of the coupling at
a fixed scale, for instance $\as(Q)$.  As long as $\vep<0$,
the running coupling vanishes for $u^2\to 0$ order-by-order in
this expansion, and the integrals all exist.  Our 
ignorance of the true behavior of $\as(u,\vep)$,
however, may be considered a signal of power
corrections \cite{BBpowerreview}, which we discuss briefly below.

In terms of $\rho^i_A$ 
the eikonal hard-scattering function simplifies to
\bea
&&\hspace{-6mm}
\ln\; \tilde \omega^i_A(N,Q) \nonumber \\[2mm]
&=&
\int_0^{Q^2} du^2\;
\frac{\rho^i_A \left (\as(u,\vep),\vep \right)}{u^2}\;
\left [ K_0\left({2Nu\over Q} \right) - \ln{Q\over u}\, \right]
\nonumber\\
&\ & \hspace{5mm} +\ 2 \ln\bar{N}\, \int_0^{Q^2} \frac{du^2}{u^2}\, 
A_i\left(\as(u,\vep)\right) \; .
\label{eq24}
\eea
The double-logarithmic structure of the exponential is
manifest in this form, with additional logarithms associated
only with the running of the QCD coupling.  Both $u$ integrals show
a (collinear) divergence at $u^2=0$.   Factorization
theorems require that these divergences
cancel between the eikonal cross section and the
collinear subtraction (the $A$ term). This implies that 
the function $\rho^i_A$ is given, up to terms that 
vanish at $\vep=0$, by the universal ``cusp" anomalous dimensions $A_i$:
\bea
\rho^i_A \left (\as(u,\vep),\vep \right) = 
2A_i\left(\as(u,\vep)\right) + F^i_A\left (\as(u,\vep),\vep \right) ,
\nonumber\\
\eea
where $F^i_A$ 
is a function that vanishes at $\vep=0$. 
However, its integral over $u$ in Eq.~(\ref{eq24}) need not vanish.

We can isolate the contribution of $F^i_A$ 
to the hard-scattering function in (\ref{eq24}) 
by an integration by parts.  For this purpose, we introduce
a new function, $D^i_A$, defined by
\bea
\int_0^{Q^2} \frac{du^2}{u^2}\; 
F^i_A\left (\as(u,\vep),\vep \right) = 
D^i_A\left (\as(Q,\vep),\vep \right)\, .
\eea
The function $D^i_A(\as,\vep)$ does not
necessarily vanish at $\vep=0$, and is related to 
$F^i_A$ by
\bea
F^i_A\left (\as(u,\vep),\vep \right) = 
\frac{\partial}{\partial \ln u^2} 
D^i_A\left (\as(u,\vep),\vep \right)\, .
\eea
For the logarithm of the eikonal hard-scattering function, we now have,
after an integration by parts, a universal expression in terms of two
conventional anomalous dimensions, $A_i$ and $D^i_A$:
\bea
\ln\; \tilde \omega^i_A(N,Q)
&=& \nonumber\\
&\ & \hspace{-25mm} \int_0^{Q^2} \frac{du^2}{u^2}\;
2 A_i\left(\as(u)\right)\; 
\left [ K_0\left({2Nu\over Q} \right) - 
\ln{Q\over \bar{N} u}\, \right]
\nonumber\\
&\ & \hspace{-30mm}
+\ \int_0^{Q^2} du^2\; 
D^i_A\left (\as(u) \right)\,
\frac{\partial}{\partial  u^2} \, 
\left [ K_0\left({2Nu\over Q} 
\right) - \ln{Q\over  u}\, \right]\, .
\nonumber\\
\label{omegawithD}
\eea
Here we have set $\vep=0$ on both sides of the equation, because 
the integrals on the right are now both finite in this limit.
The first ($A$) term generates the leading double logarithms, because 
for $u \gg Q/N$ the Bessel function $K_0(2Nu/Q)$
vanishes rapidly, leaving 
a $\ln(\bar{N}u/Q)/u^2$ behavior.   For 
the second ($D_A^i$) term, the 
derivative of the expression in square brackets
 is again power-suppressed when $u \ll Q/N$,
but lacks the logarithmic enhancement for larger $u$, where
it behaves simply as $1/u^2$,
\bea
\frac{\partial}{\partial u^2}\, \left [ K_0\left({2Nu\over Q} \right) - 
\ln{Q\over  u}\, \right] &\ &
\nonumber\\
&\ & \hspace{-30mm} \sim - \frac{N^2}{Q^2}\,
\left( \ln \frac{\bar{N}u}{Q} - \frac{1}{2}\right)\, 
 \, ,\quad \, \frac{Nu}{Q} \ll 1\, ,
\nonumber\\
&\ & \hspace{-30mm} \sim   \frac{1}{2u^2} \, , \quad\quad  \frac{Nu}{Q} \gg 1\, .
 \eea
  Because the function  $D^i_A$ receives no contributions from a single-gluon final 
state \cite{LSV},  the second term in
Eq.\ (\ref{omegawithD}) begins at next-to-next-to-leading 
logarithm in the moment variable. This term generalizes the 
$D$-terms found in Drell-Yan threshold resummation~\cite{dyresum1,dyresum2,Drefs}.  
We note again that this result is found by consistently using the
coupling in $4-2\vep$ dimensions, with $\vep<0$.  
The final expression, however, is finite for $\vep\to 0$, as
it must be in the short-distance function $\tilde \omega^i_A(N,Q)$.

We also see explicitly from (\ref{omegawithD}) that
the expression in Eq.\ (\ref{nextA2}) vanishes for $N\to 0$,
 which can be thought of as
a normalization condition for virtual corrections.  
This confirms that the resummed short-distance function 
has been constructed to consist of plus distributions only.  It may, however,
be corrected by $N$-independent constants \cite{Eynck:2003fn}.  

\medskip

\section{Discussion \label{sec6}}

We have shown that the full formalism for threshold resummation
can be extended from the Drell-Yan cross section to its
crossed relatives in single-inclusive deep-inelastic scattering
and double-inclusive annihilation.   The
hard-scattering functions for all of
these processes can now be expressed in
terms of exponentiated integrals of eikonal web functions,
related by crossing eikonal lines.  We have also shown
how process-dependent nonleading logarithms
arise naturally in the eikonal formalism and that, within
this set of diagrams, the relevant ($D^i_A$) anomalous dimensions
are all closely related.   These results 
extend the set of observables that can be 
analyzed to all orders and logarithms in terms of
a limited set of anomalous dimensions, and
should facilitate the investigation of the relationship
between perturbative and nonperturbative dynamics
in these and related hard-scattering processes.
Here we shall make a few observations on possible applications.

\subsection{Joint resummation}

Following the analysis of 
Ref.\ \cite{LSV}, the threshold-resummed
hard scattering functions above can all be extended 
to joint resummation.
At leading power in $N$, the double transform to Mellin moment 
and impact parameter space in joint resummation,
is derived by simply inserting the factor $\exp[i{\bf b}\cdot {\bf k}_T]$ 
in the integral over final-state momenta in Eq.\ (\ref{nextA2}):   
\begin{widetext}
\beeq
\ln\; \tilde \omega^i_A(N,Q,b)
&=&
2\, \int_0^{Q^2} du^2\; \Bigg\{\int_{k_T^2\leq u^2} \, 
{d^{2-2\epsilon}k_T\over 
[2\pi^{1-\varepsilon}/\Gamma(1-\varepsilon)]}\,
W^i_A\left(k_T^2,u^2,u^2,\alpha_s(u,\varepsilon),
\varepsilon\right) \nonumber \\
& & \hspace{5mm} \times
\left[\, {\rm e}^{i{\bf b}\cdot {\bf k}_T}
K_0\left({2Nu\over Q} \right) - \ln{Q\over
u}\, \right]\
+ {2\over u^2}\ln \bar N\, 
A_i\left(\as(u,\vep)\right) \Bigg\} \, .
\label{nextAu2}
\eeeq
\end{widetext}
That is, at leading power of $N$,
all logarithms in $b$ are controlled
by the same web functions as in threshold
resummation.  The extension to $b$-dependence
at nonleading powers of $N$ was investigated
in \cite{KSV} for Drell-Yan processes.  These
considerations may be extended to the crossed
reactions at observed transverse momenta in a straightforward manner.


\subsection{Power corrections}

In the resummed expressions of
Eq.\ (\ref{omegawithD}), as in resummations of many 
other physically-relevant quantities, perturbative non-convergence
arises  from the integral over a momentum 
scale $u$ that can be identified with the argument of
the strong coupling, $\as(u)$.  In cases where $u=0$
is the endpoint of the integral,
re-expanding the running coupling $\as(u)$ in terms of any
coupling at fixed scale, say, $\as(Q)$, leads to integrals that 
are finite for infrared-safe quantities, but which grow factorially
with the order.   Generally speaking, the presence
of the form $(1/Q)^p\int_0^Q d\mu \mu^{p-1}\ f(\as(\mu))$
is taken to imply  the presence of a $1/Q^p$ power
correction~\cite{BBpowerreview,DSeventreview,Laterevent}.  A conjecture of universality is 
natural when the function $f(\as)$ is the same 
for different processes.   Expanding the $K_0$ function,
we find for each of these observables an expansion in
even powers (only) of $N/Q$ and $bQ$.

Many investigations of power corrections
have been carried out in this fashion~\cite{BBpowerreview,DSeventreview,Laterevent}, normally on
the basis of cross sections that have been resummed
to next-to-leading logarithm.   
It is interesting, therefore, to make a similar
analysis in these cases, where the resummation
has been carried out in principle for all logarithms.

This reasoning would seem to imply
identical power corrections from 
the  $A_i$ term in Eq.\ (\ref{omegawithD}),
which generates leading and next-to-leading logarithms
in perturbation theory,
since this anomalous dimension is the same in each of the
cross sections.  
The  anomalous dimensions $D^i_A$,
on the other hand, may vary between these processes, although
only through the dependence of the web functions
$W^i_A$ on the signs of the invariants $\beta\cdot k$, $\beta' 
\cdot k$, as mentioned above. It will clearly be of interest to investigate
further the possible influence of the $D$ terms on power
corrections in these cross sections, where the perturbative
structure is relatively simple. 

\subsection{Phenomenological applications and generalizations}

We close with the observation that the cross sections
resummed in this paper can in principle be compared
directly with data from deep-inelastic and leptonic annihilation
processes over a wide range of energy scales.  
It may also be possible to generalize this analysis to a number  of
observables in hadron-hadron scattering, for example
dihadron cross sections.  We leave these applications
and generalizations to future work.

\vspace*{5mm}
\subsection*{Acknowledgements}

The work of G.S.\ was supported in part
by the National Science Foundation, grants PHY-0354776,
and PHY-0354822. W.V.\ is grateful to RIKEN, BNL
and the U.S.\ Department of Energy (contract number DE-AC02-98CH10886) for
providing the facilities essential for the completion of part of 
this work.

\end{document}